# An approach towards a perfect thermal diffuser

Krishna P. Vemuri and Prabhakar R. Bandaru*,

*Department of Mechanical Engineering, University of California, San Diego,*

*La Jolla, CA 92093-0411*

**Abstract:** A method for the most efficient removal of heat, through an anisotropic composite, is proposed. It is shown that a rational placement of constituent materials, in the radial and the azimuthal variation, at a given point in the composite yields a uniform temperature distribution in spherical diffusers, accompanied by a very significant reduction of the source temperature, in principle, to infinitesimally above the ambient temperature, and underlies the design of a *perfect* thermal diffuser. Orders of magnitude enhanced performance, compared to that obtained through the use of a diffuser constituted from a single material with isotropic thermal conductivity has been observed and the analytical principles underlying the design were validated through extensive computational simulations.

**Main Text:**

The rapid dissipation of thermal energy from a heat source as well as an accurate control of the path of heat transfer is necessary for a wide variety of applications, and has been considered under the scope of efficient thermal management(*1, 2*). At the very outset, it may be thought that the optimal method of dissipating thermal energy would involve omnidirectional transfer and could be mediated either through spherical or cylindrical symmetry, for point and line sources of heat, respectively. However, as we will show, materials design considering isotropic thermal conductivity is inefficient (*1*). Additionally, spherical/cylindrical thermal energy diffusers based on such considerations yield nonlinear temperature distribution and concomitant non-uniform heat flux. The staggering of the temperature throughout the volume of the heat-dissipating element in such cases may also result in the creation of undesirable *hot spots*(*3*). As a solution to the above shortcomings, in this work, we report on the fundamental notion of the optimal materials arrangement necessary for the *most* efficient removal of heat, in which a linear temperature profile along with the desirable characteristic of isotropic heat transfer would be obtained. We will show that, in the steady state, our design allows for a significant reduction of the source temperature.

We examine a typical situation where a given steady-state heat flux ($q_{in}$), produced by a thermal energy source (which releases heat uniformly in all directions), needs to be dissipated to the ambient (at a temperature of $T_{amb}$). For this purpose, a thermal diffuser of a certain shape/geometry needs to be placed between the heat source and the ambient. If any rectangular geometry, comprising a material with isotropic and constant thermal conductivity ($\kappa$), were chosen, the azimuthally asymmetric geometry results in a non-uniform temperature gradient and anisotropic heat transfer. In comparison, isotropic heat transfer with *unequally* spaced isotherms (along the heat flux direction) would be obtained in circular geometries with the constant thermal conductivity. This results in an undesired staggering of isotherms near the heat source: Fig. 1(A). A more desired diffuser configuration, wherein the isotherms would be equally spaced yielding a uniform temperature distribution, is indicated in Fig. 1(B).

The design of the desired heat diffuser proceeds through a conformal mapping(*4, 5*) from the rectangular geometry - shown in Fig. 2(A) to a circular configuration – shown in Fig. 2 (B), so as to obtain a linear temperature gradient concomitant with isotropic heat transfer. The space spanned in the rectangular geometry: **U** = (x, y, z) is related to the space in the circular geometry: **V** = (X, Y, Z) through a relation of the form:

$$X = y Sin\left(\frac{x}{R}\right), Y = y Cos\left(\frac{x}{R}\right), Z = z \tag{1}$$

The mapping of the (*x, y, z*) space to (*X, Y, Z*) space, indicated through Eqn. 1, is conformal as it preserves the parallelism of the isotherms in the two spaces. In these relations, *R* refers to a mean radius pertinent to the semi-circular strip of width: *2W* (the inner and the outer radius of the circular strip, in Fig. 2(B), are $R_i$ and $R_o$, and $R = [R_i + R_o]/2$ and $W = [R_o - R_i]/2$). For a correspondence to the (*r, θ, z*) system in this figure, $x = R\, Tan^{-1}\,(X/Y) = R\phi$, and $y = \sqrt{X^2 + Y^2} = r$. Symmetry considerations allow the probing of a semi-circular half space and a two-dimensional layout. The thermal conductivity of one coordinate system (say, the rectangular system, where the thermal conductivity is denoted through $\kappa^{rect}$) could be related to another (say, the circular system with thermal conductivity: $\kappa^c$) through a transformation(*6–8*) described through: $\kappa^c = (J\kappa^{rect}J^T)/\det(J)$, where *J* is the Jacobian for the considered transformation, $J^T$ is the transpose, and det (*J*) is the determinant of the Jacobian. It can then be derived (see Section I

of the Supplementary Materials for details) that the thermal conductivity of the semi-circular

diffuser in Fig. 2(B), would be a function of the distance from the heat flux source ($r$), through:

$$\kappa^c(r,q,z) = \begin{pmatrix} \kappa_r^c & 0 & 0 \\ 0 & \kappa_q^c & 0 \\ 0 & 0 & \kappa_z^c \end{pmatrix} = \begin{pmatrix} \dfrac{R}{r} & 0 & 0 \\ 0 & \dfrac{r}{R} & 0 \\ 0 & 0 & \dfrac{R}{r} \end{pmatrix} \kappa \; ; \; \begin{array}{l} R - W \leq r \leq R + W \\ 0 \leq q \leq 2f \end{array} \quad (2)$$

The physical implication, of Eqn. (2), is that the regularization of the heat transport, to obtain a uniform temperature distribution and isotropic heat transfer, requires the modification of matter following the path of heat transfer. Consider, for instance, the thermal conductivity in the radial direction $\kappa_r^c$. At $r < R$ (/ $r > R$), $\kappa_r^c$ ($= \dfrac{R}{r}\kappa$) would be larger (/smaller) than the nominal isotropic thermal conductivity $\kappa$, implying that the resultant distance between isotherms in the anisotropic diffuser: Fig. 1(B), would be farther (/closer) than its isotropic counterpart: Fig. 1(A). Such imposed variation, yielding a respective expansion (/contraction) of the isotherm spacing close to (/further away) from the heat source in the spherical geometry, would help achieve a linear temperature profile along the heat flux direction. The spatial change of the $\kappa^c(r)$, indicated in Fig. 3(A), would indicate that the thermal conductivity of the semi-circular heat diffuser would be identical to that of the isotropic thermal conductivity only at $r = R$, and det ($J$) = 1 for the semi-circular geometry with respect to the rectangular geometry. While the through thickness variation in the diffuser, through the $\kappa_z^c$ may be similarly interpreted, the thermal conductivity change in the azimuthal direction $\kappa_q^c(r)$, follows an inverse dependence on $r$. As such unusual anisotropy cannot be found in nature, at any particular $r$, the radial/perpendicular and azimuthal variation of the thermal conductivity must be considered through suitable and specific material placement(*7*, *9*) for obtaining uniform temperature distribution.

In this regard, we have experimentally demonstrated in previous work(*6*) that a layered composite of just two disparate materials (say, of thermal conductivity: $\kappa_1$ and $\kappa_2$) and of thickness ($l$) would be sufficient to obtain any desired anisotropic thermal conductivity at a given $r$ (see Section II of the Supplementary Materials). Moreover, the influence of any particular value of $l$ on the propagation of heat flux can be eliminated through the creation of an *effective thermal medium* (ETM)(*10*). Following Fig. 3(A), at $r < R$, the thermal conductivity in the radial (/perpendicular) direction: $k_r^c$ (/ $k_z^c$) would be larger than the $k_q^c$, implying an orientation of elements parallel to the heat flux, while for $r > R$, *i.e.*, when $k_r^c$ (/ $k_z^c$) is smaller compared to the $k_q^c$, the orientation of the elements would be perpendicular to the heat flux. Essentially, the size of a composite element (yielding the desired $\kappa^c$) should be such that a constant temperature gradient across the element could be assumed, in accordance with the tenets of an ETM. We will show later, through computational simulations, that the related arrangement of materials indeed ensures uniform heat flow and a linear temperature gradient. The facilitating methodology involves a metamaterial architecture(*11–14*), constituted from individual *thermal meta-atoms* at a given $r$, and with spatially varying values of the thermal conductivity (*i.e.*, through $k_r^c$, $k_z^c$, and $k_q^c$). Such $r$ dependent variation yields an *anisotropic* (*10*) character to the diffuser.

We discuss next the temperature variation in the thermal diffuser employing such spatial conductivity modulation. Through consideration of the heat flux continuity in (a) an *isotropic* cylindrical diffuser: Fig. 1(A), *vis-à-vis,* (b) an *anisotropic* cylindrical diffuser: Fig. 1(B), and a convective heat transfer coefficient $h$ between the outer radial surface $r = R_o$ and the ambient (at

$T_{amb}$), it can be derived (see Section III of the Supplementary Materials) that the source temperature (at $r = R_i$), in the *isotropic* and the *anisotropic* case, would be respectively:

$$T^{iso}_{r=R_i} = q_{in}\left[\frac{R_i}{k}\ln\left(\frac{R_o}{R_i}\right) + \frac{1}{h}\left(\frac{R_i}{R_o}\right)\right] + T_{amb} \tag{3a}$$

$$T^{aniso}_{r=R_i} = q_{in}\left[\frac{R_i}{kR}(R_o - R_i) + \frac{1}{h}\left(\frac{R_i}{R_o}\right)\right] + T_{amb} \tag{3b}$$

While Eqn. 3(a) predicts a logarithmic temperature variation, it is apparent from Eqn. 3(b) that a linear temperature profile is obtained through the use of an anisotropic materials architecture: Fig. 3(B). The latter attribute now yields a lower temperature at the source ($r = R_i$) by an amount, $\Delta T_{r=Ri}$, obtained by subtracting (3b) from (3a), and is:

$$\Delta T_{r=R_i} = \frac{q_{in}R_i}{\kappa}\left[\ln\left(\frac{R_o}{R_i}\right) - 2\left(\frac{R_o - R_i}{R_o + R_i}\right)\right] \tag{4}$$

One way to understand Eqn. (4), from a physical perspective, involves considering the first term: $\frac{q_{in}R_i}{\kappa}$, as the thermal energy input, while the multiplying bracketed term represents the reduced temperature due to induced anisotropy. The resulting plots of $\Delta T_{r=R_i}$ as a function of the mean radius ($R = \frac{1}{2}[R_i + R_o]$) and the width of the diffuser ($W = \frac{1}{2}[R_i - R_o]$) are shown in Fig. 4. The constraint of $R_i > 0$, mandates that $R > W$ and marks the beginning of the curve. At large $R$ (/small $W$), the anisotropic composite would degenerate to an isotropic case, implying a very small $\Delta T_{r=R_i}$. We also observe that at increasingly larger $R$, with a fixed $W$, the logarithm in Eqn. (4) could be linearized yielding a correspondingly diminished temperature difference. The peak in the $\Delta T_{r=R_i}$ corresponds to an $r$ where the enhanced thermal diffusion due to the induced anisotropy would be balanced by the increasing heat input. A very significant reduction of the source temperature can be obtained, in principle to infinitesimally above the ambient temperature, and lays the basis for a *perfect* thermal diffuser.

Computational simulations were performed to validate our analytical derivations and the principles underlying the perfect thermal diffuser design: Fig. 5(A). We consider a diffuser needing to dissipate a $q_{in}$, of $2 \times 10^6$ W/m$^2$, as may occur in heat assisted magnetic recording (HAMR) (*15*, *16*), with $R$ = 2.5 cm and $W$ = 2 cm. The corresponding anisotropic thermal conductivity of the diffuser and placement of the material, per Eqn. (2), was derived (also see Section IV of the Supplementary Materials) assuming a nominal isotropic $\kappa$ = 60W/mK. A comparison of the analytically predicted temperature variation from Eqn. 3b, with the computational results is illustrated in Fig. 5(B), *cf.* Fig. 3(B). The temperature variation was found to be linear in excellent accord with the predictions, and has been achieved through the use of relatively few layers. As predicted, a remarkable reduction in the source temperature (following Eqn. 4) by ~ 100 K in the engineered composite, is a highlight of our design and indicates the validity of our approach as a proposal for a new type of a thermal management technique.

In summary, we have implemented a simultaneous variation of the radial and the azimuthal variation of the thermal conductivity at a given point in an anisotropic material and have hence demonstrated the path towards creating a perfect thermal diffuser. The significant lowering of the source temperature in such a diffuser, compared to that obtained through the use of a single material with isotropic thermal conductivity, would find many applications. Our work also exemplifies the utility of anisotropic architectures. As heat transfer is fundamentally anisotropic, as (a) in thermal conduction, *e.g.,* due to phonon dispersion(*17*, *18*), (b) in convection, *e.g.,* due to buoyancy effects(*19*), and (c) in radiative heat transfer, *e.g.,* due to view factor related effects(*20*), the regularization of diffusive heat transfer through the use of anisotropic composites may indeed be appropriate.

**Acknowledgements:**

The authors are grateful for support from the Defense Advanced Research Projects Agency (DARPA: W911NF-15-2-0122), National Science Foundation (NSF: CMMI 1246800), Qualcomm Inc., and interactions with Prof. T. Yang, and F. Canbazoglu.

**Figures:**

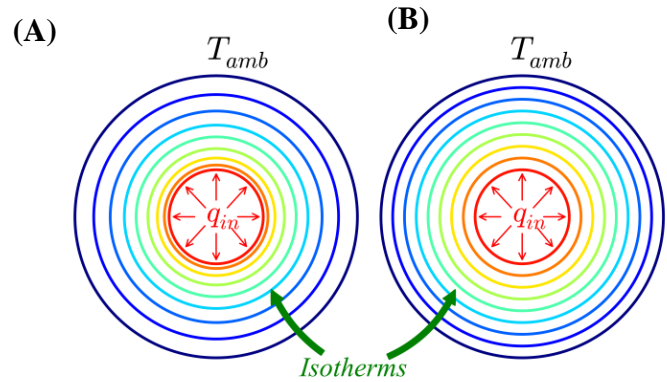

**Fig. 1.** A comparison of diffusional heat flow in spherical geometry, in a thermal diffuser, constituted from (**A**) a material with isotropic and a constant thermal conductivity, with unequally spaced isotherms, and (**B**) an ideal configuration with uniformly spaced isotherms and lower source temperature. The latter is feasible through the optimized arrangement of the constituent materials in a composite thermal diffuser.

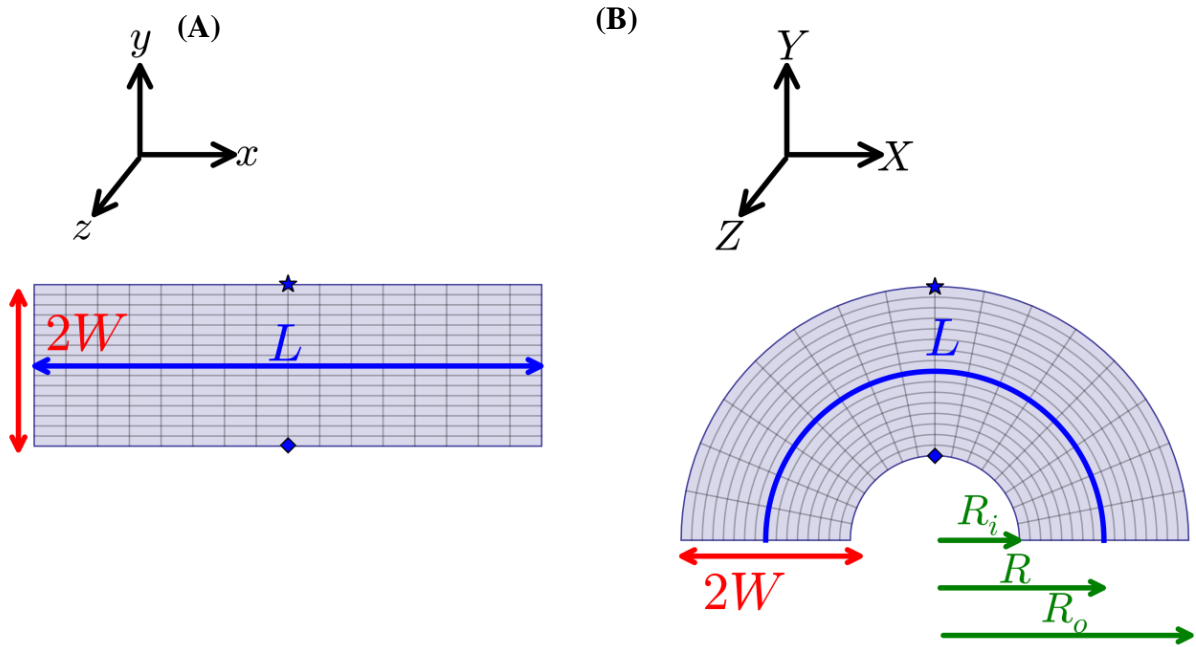

**Fig. 2.** Isotropic heat transfer concomitant with uniform temperature distribution is obtained through the conformal mapping of the thermal conductivity of a **(A)** rectangular block of length $L$ and width $2W$ on to a **(B)** semi-circle, with inner radius $R_i$, outer radius $R_o$ and width $2W$. Both the geometries in (A) and (B) have the same out of plane thickness. Two representative correspondent points (indicated, for example, by the ◆ and ▫) are indicated.

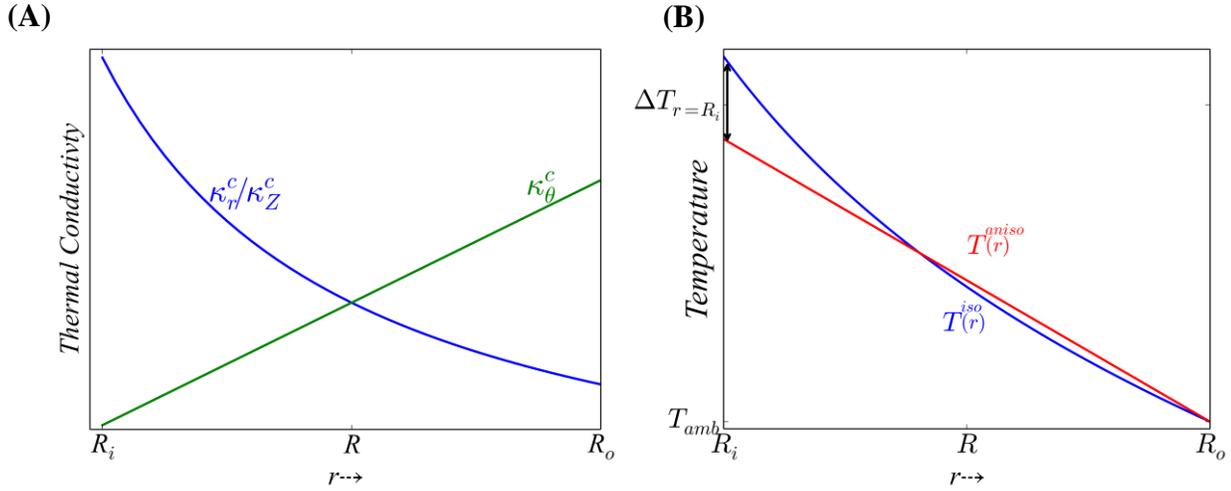

**Fig. 3.** **(A)** The variation of the *in plane* radial ($\kappa_r^c$), and azimuthal ($\kappa_\theta^c$) thermal conductivity, along with the *cross-plane* thermal conductivity ($\kappa_Z^c$) with radial distance (*r*) from the heat source. **(B)** While the radial variation of the temperature, *i.e.*, $T^{iso}(r)$, is nonlinear for a diffuser constituted from isotropic material, a linear temperature gradient together with a resultant lower source temperature may be obtained through the use of an anisotropic material (the temperature variation of which is depicted through $T^{aniso}(r)$.

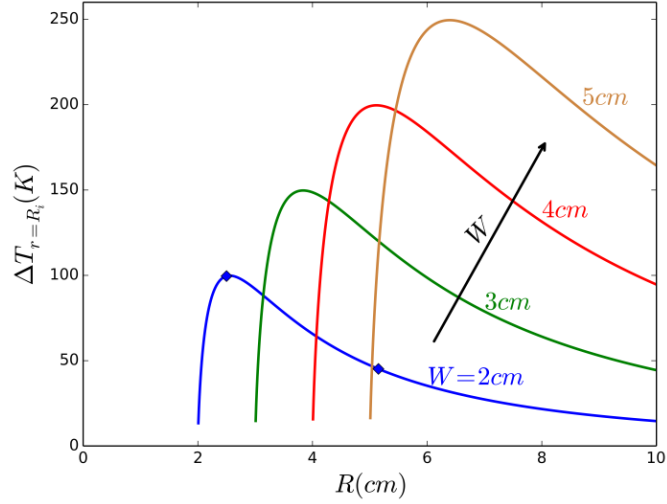

**Fig. 4.** The variation of the difference in temperature, at the source ($r=R_i$): $\Delta T_{r=R_i}$, between using a thermal diffuser constituted from a single material with *isotropic* thermal conductivity and the desired *anisotropically* architected material, with the mean radius ($R = \frac{1}{2}[R_i + R_o]$) and the width of the diffuser ($2W = [R_i - R_o]$), per Eqn. (4). The marked points (♦) indicate results from the computational simulations.

**(A)**  **(B)**

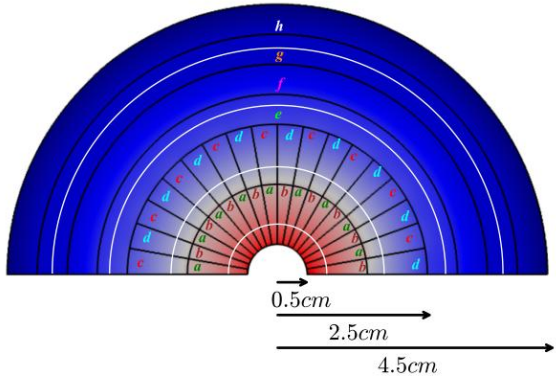 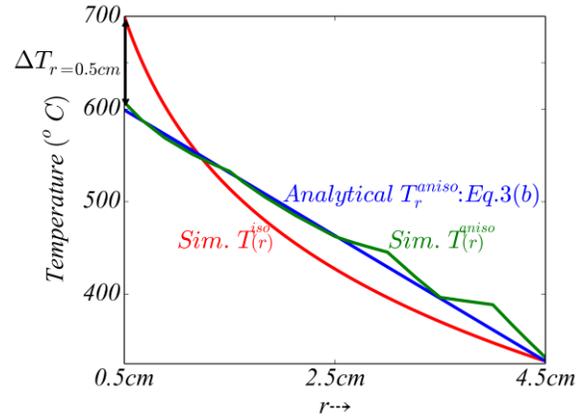

**Fig. 5.** **(A)** Computational simulations indicating the notion of a perfect thermal diffuser, for an input heat flux $q_{in} = 2\times10^6$ W/m$^2$. A uniform temperature distribution in the composite was obtained through the use of an anisotropic arrangement (see Table S1 of the Supplementary Materials for the details on the arrangement scheme), composed of materials of constant and isotropic thermal conductivity. The symbols refer to the nominal thermal conductivity of the used isotropic materials, *i.e.*, *a* (= 282 W/mK), *b* (= 12 W/mK), *c* (= 118 W/mK), *d* (= 29 W/mK), *e* (= 110 W/mK), *f* (= 31 W/mK), *g* (= 169 W/mK), and *h* (= 20 W/mK). **(B)** The resulting temperature variation, for the isotropic case: $T^{iso}(r)$, and the anisotropic case: $T^{aniso}(r)$, determined from the simulations (*Sim.*), was found to be in excellent accord with the analytical (*Analytical*) relation derived in Eqn. 3(b).

# Supplementary Materials

# An approach towards a perfect thermal diffuser

Krishna P. Vemuri and Prabhakar R. Bandaru,

*Department of Mechanical Engineering, University of California, San Diego,*

*La Jolla, CA 92093-0411*

# I. Obtaining a uniform temperature distribution (rectangular space) and isotropic heat flow (spherical space), through mutual mapping

The mapping transformation involves the mapping(*1*) of a rectangular block, formed from an isotropic material of thermal conductivity: $\kappa$, and of length: $L$ Figure S1 (a), onto a cylindrical sector: Figure S1 (b), The correspondent mapping between a space spanned in rectangular geometry, *i.e.*, $\mathbf{U} = (x, y, z)$ to that of a space spanned by cylindrical geometry, *i.e.*, $\mathbf{V} = (X, Y, Z)$, accomplishes the incorporation of features relevant to a uniform temperature distribution coupled with isotropic diffusion.

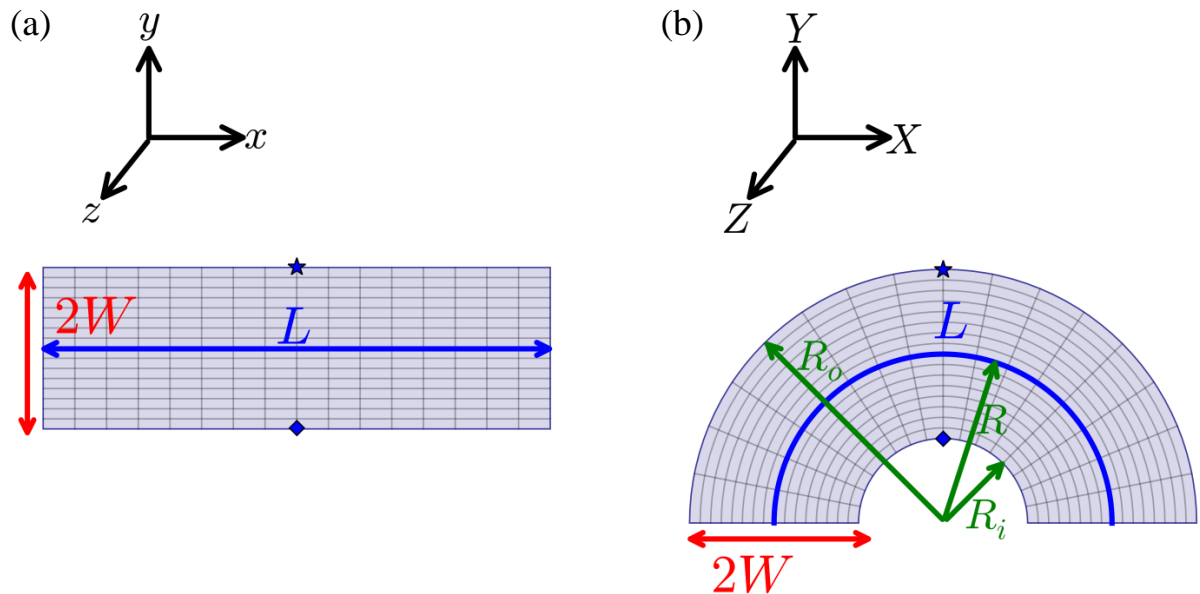

**Figure S1** Schematic mapping of a **(a)** rectangular block of length $L$, width $2W$ and thickness $t$ mapped onto **(b)** a cylindrical sector with central angle $2\theta$, inner radius $L/\theta - W$, outer radius $L/\theta + W$ and thickness $t$. Two representative correspondent points (indicated, for example, by the ◆ and ☐) are indicated. Note that, as depicted, $\theta = \pi/2$, $R_i = L/\theta - W$, $R_o = L/\theta + W$, $R_o - R_i = 2W$.

The mapping of the (x, y, z) space to (X, Y, Z) space, indicated in Eqn. S1, below, is conformal and preserves the parallelism of the isotherms in the two spaces. The mutual geometric relationships are represented through:

$$X = y\mathrm{Sin}\left(\frac{x}{R}\right); Y = y\mathrm{Cos}\left(\frac{x}{R}\right); Z = z \tag{S1a}$$

or, equivalently

$$x = R\mathrm{Tan}^{-1}\left(\frac{X}{Y}\right), \ y = \sqrt{X^2 + Y^2}, z = Z \tag{S1b}$$

The Jacobian (J), for the transformation, is given by (2–4) $J_{U \to V} = \frac{\partial(X,Y,Z)}{\partial(x,y,z)}$:

$$J_{U \to V} = \begin{pmatrix} \frac{y}{R}\mathrm{Cos}\left(\frac{x}{R}\right) & \mathrm{Sin}\left(\frac{x}{R}\right) & 0 \\ \frac{-y}{R}\mathrm{Sin}\left(\frac{x}{R}\right) & \mathrm{Cos}\left(\frac{x}{R}\right) & 0 \\ 0 & 0 & 1 \end{pmatrix} \tag{S2}$$

An alternate representation related to **Eqn. (S2)** involves the use of a cylindrical coordinate system (from Eqn. S1b), and with reference to the rectangular coordinate system at the top of Figure S1(b), and uses:

$$\begin{aligned} x &= R\mathrm{Tan}^{-1}(\frac{X}{Y}) = Rq \\ y &= \sqrt{X^2 + Y^2} = r \\ z &= Z \end{aligned}$$

(S3)

The modification of the thermal conductivity, due to the transformation given by Eqn. (S1), is then obtained through:

$$k^{rect}(r,q,Z) = \frac{J_{U \to V} k J_{U \to V}^T}{Det(J_{U \to V})} = \begin{pmatrix} \frac{r}{R}\text{Cos}^2(q) + \frac{R}{r}\text{Sin}^2(q) & \frac{1}{2Rr}(R^2 - r^2)\text{Sin}(2q) & 0 \\ \frac{1}{2Rr}(R^2 - r^2)\text{Sin}(2q) & \frac{r}{R}\text{Sin}^2(q) + \frac{R}{r}\text{Cos}^2(q) & 0 \\ 0 & 0 & \frac{R}{r} \end{pmatrix} k \quad \text{(S4)}$$

The equivalent thermal conductivity in a rectangular (*X, Y, Z*) coordinate system is done through a rotation by an angle $\theta$ about the Z-axis, as follows:

$$k^{circ}(X,Y,Z) = \frac{J_2 k^c(r,q,Z) J_2^T}{Det(J_2)} = \begin{pmatrix} k_r^c & 0 & 0 \\ 0 & k_q^c & 0 \\ 0 & 0 & k_z^c \end{pmatrix} = \begin{pmatrix} \frac{R}{r} & 0 & 0 \\ 0 & \frac{r}{R} & 0 \\ 0 & 0 & \frac{R}{r} \end{pmatrix} k \quad \text{(S5)}$$

where, $J_2 = \begin{pmatrix} \text{Cos}\theta & -\text{Sin}\theta & 0 \\ \text{Sin}\theta & \text{Cos}\theta & 0 \\ 0 & 0 & 1 \end{pmatrix}$

Such a form has been indicated as Eqn (2) of the main text and the schematic variation of conductivity ($k_r^c$, $k_z^c$, and $k_q^c$, has been plotted in **Fig. 3(A)**, in the main text.

## II. Constructing the spatial anisotropy represented in Eqn. (2)/(S5) using isotropic materials

The spatial anisotropy represented through the relation in Eqn. (2) of the main text or (S5) in Section I of the Supplementary information, can be effectively reduced to a two-dimensional form, in the ($r$, $\theta$) system, as $\kappa_r^c = \kappa_z^c = \dfrac{R}{r}$. Also, it is noted from (S5) and Figure 3(a), that

$$\begin{aligned} \kappa_r^c > \kappa_\theta^c \quad &; \quad R_i \leq r < R \\ \kappa_r^c = \kappa_\theta^c \quad &; \quad r = R \\ \kappa_r^c < \kappa_\theta^c \quad &; \quad R < r \leq R_o \end{aligned} \tag{S6}$$

A composite thermal conductivity satisfying the above conditions can be obtained through the arrangement of two layers, with isotropic thermal conductivity values of $\kappa_1$ and $\kappa_2$. In our previous work(2, 7), it was shown that the effective thermal conductivity of such layers arranged in parallel, i.e., $\kappa_p$ was larger than that of layers arranged in series, parallel, i.e., $\kappa_s$.

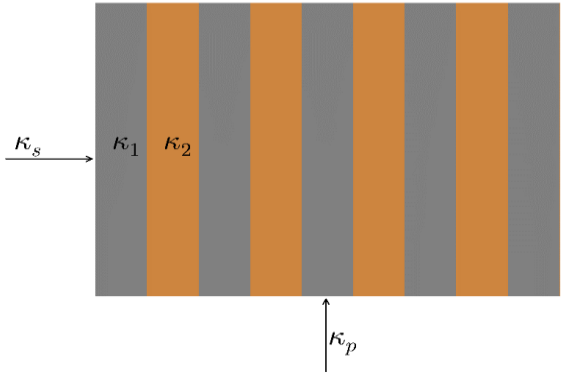

**Figure S2** The alternate stacking of two layers, with isotropic thermal conductivity values of $\kappa_1$ and $\kappa_2$, respectively, can yield an effective composite thermal conductivity measured in series, i.e., $\kappa_s$, and another *larger* value for parallel measurement, i.e., $\kappa_p$.

For example, with reference to Figure S2, it may be derived that:

$$\kappa_p = \frac{\kappa_1 + \kappa_2}{2} \;,\; \kappa_s = \frac{2\kappa_1\kappa_2}{\kappa_1 + \kappa_2} \tag{S7a}$$

Then,

$$\kappa_1 = \kappa_p + \sqrt{\kappa_p^2 - \kappa_p \kappa_s} \;,\; \kappa_2 = \kappa_p - \sqrt{\kappa_p^2 - \kappa_p \kappa_s} \tag{S7b}$$

Consequently, for $r < R$, the orientation of layered elements would need to be parallel to the heat flux, with a correspondence of the $k_r^c$ ($/k_z^c$) to $\kappa_p$ and with $k_q^c$ being related to the $\kappa_s$. However, for $r > R$, the orientation of layered elements would need to be arranged *in series* with the heat flux, with a correspondence of the $k_r^c$ ($/k_z^c$) to $\kappa_s$ and $k_q^c$ related to $\kappa_p$. This would then imply a spatially varying $\kappa_1$ $(r)$ and $\kappa_2$ $(r)$.

**III. Temperature variation in thermal diffuser employing spatial conductivity modulation**

We consider cylindrical diffusers, of the type shown in Fig. 2(A) – in the main text, or Figure S1 (b), with an input heat flux $q_{in}$ at $r = R_i$ ($= R-W$) placed in an ambient (at a temperature $= T_{amb}$). A given convective heat transfer coefficient $h$ between the outer surface at $r = R_o$ ($= R+W$) and the ambient is assumed (5, 6). The resulting temperature, at the outer surface, for the cylindrical diffusers, with (a) homogeneous and isotropic material ($T_{R_O}^{iso}$), and (b) with anisotropic and spatial modulation of the constituents ($T_{R_O}^{aniso}$), are then derived.

From thermal energy conservation,

$$q_{in} R_i t = h R_o t \left( T_{R_o}^{iso} - T_{amb} \right) = h R_o t \left( T_{R_o}^{aniso} - T_{amb} \right)$$

$$T_{R_O}^{iso} = T_{R_O}^{aniso} = T_\infty + \frac{q_{in}}{h}\left(\frac{R_i}{R_o}\right) \tag{S8}$$

(a) *Temperature $T_r^{iso}$ in an isotropic cylindrical diffuser*

From heat flux continuity, for an isotropic cylindrical diffuser $\kappa$,

$$q_{in} R_i t = -k r t \frac{\partial T(r)}{\partial r} \tag{S9}$$

Integrating Eqn. (S7) between $r = r$ (at a temperature, $T = T_r^{iso}$) and $r = R_o$ (at $T = T_{R_o}^{iso}$) and using the relation in (S6) for the boundary conditions, we get,

$$T_r^{iso} = \frac{q_{in} R_i}{k} \ln\left(\frac{R_o}{r}\right) + T_{amb} + \frac{q_{in}}{h}\left(\frac{R_i}{R_o}\right) \tag{S10}$$

Consequently, $\qquad T_{r=R_i}^{iso} = q_{in} \left[ \frac{R_i}{k} \ln\left(\frac{R_o}{R_i}\right) + \frac{1}{h}\left(\frac{R_i}{R_o}\right) \right] + T_{amb} \tag{S11}$

This is **Eqn. (3a)** in the main text.

(b) *Temperature $T_r^{aniso}$ in an anisotropic cylindrical diffuser*

From heat flux continuity, for an anisotropic cylindrical diffuser, with spatially varying thermal conductivity, given by $k^c(r,q,Z) = k_r^c = \frac{R}{r}k$, the heat conservation equation yields:

$$q_{in} R_i t = -k_r^c rt \frac{\partial T(r)}{\partial r} \tag{S12}$$

Integrating Eqn. (S12) between $r = r$ (at $T = T_r^{aniso}$) and $r = R_o$ (at $T = T_{R_O}^{aniso}$) and using the boundary conditions represented in (S8), we get,

$$T_r^{aniso} = \frac{q_{in} R_i}{kR}(R_o - r) + T_{amb} + \frac{q_{in}}{h}\left(\frac{R_i}{R_o}\right) \tag{S13}$$

Then,

$$T_{r=R_i}^{aniso} = q_{in}\left[\frac{R_i}{kR}(R_o - R_i) + \frac{1}{h}\left(\frac{R_i}{R_o}\right)\right] + T_{amb} \tag{S14}$$

This is **Eqn. (3b)** in the main text.

**IV. The arrangement of materials used for simulation in Fig. 5**

We indicate a representative material arrangement underlying the computational simulations needed to validate our analytical derivations and the principles underlying the perfect thermal diffuser design. Following Fig. 2(B), we use representative $R_i = 0.5$ cm, $R_o = 4.5$ cm (implying $R = 2.5$ cm and $W = 2$ cm), and $L = 2.5\,\pi$ cm, with $\kappa = 60$ W/mK.

From Eqn. (2)/ Eqn. S5, we obtain:

$$k^{circ}(X,Y,Z) = \begin{pmatrix} k_r^c & 0 & 0 \\ 0 & k_q^c & 0 \\ 0 & 0 & k_z^c \end{pmatrix} = \begin{pmatrix} \dfrac{R}{r} & 0 & 0 \\ 0 & \dfrac{r}{R} & 0 \\ 0 & 0 & \dfrac{R}{r} \end{pmatrix} k = \begin{pmatrix} \dfrac{150}{r} & 0 & 0 \\ 0 & 24r & 0 \\ 0 & 0 & \dfrac{150}{r} \end{pmatrix} \quad (S15)$$

At any given $r$, the thermal conductivity mimicking the spatial anisotropy may be determined using the above relation. We approximately sub-divide the diffuser into four regions of radial thickness 1 cm. The $k_r^c$ ($/k_z^c$) and $k_q^c$ is subsequently computed from (S15), at any given $r$ (taken for example, to be at the center of the individual regions, *e.g.*, for the first region spanning $r = 0.5$ cm and $r = 1.5$ cm, the *effective* $r = 1$ cm. Following the rationale presented in Section III of the Supplementary Materials, for $r < R$, the $k_r^c$ ($/k_z^c$) is corresponded to a $\kappa_p$ and the $k_q^c$ is related to the $\kappa_s$. Consequently, the appropriate $\kappa_1$ and $\kappa_2$ is obtained through using Eqn. (S7b). Such details for all the four sections of the anisotropic diffuser are depicted in Table I, on the next page.

**Table S1** The corresponding arrangement of materials in Fig. 5(A). It is discussed in the paper as to how the arrangement of isotropic materials in an anisotropic manner could yield uniform temperature distribution along with a lower source temperature. Following Fig. 2(B), $R_i = 0.5$ cm, $R_o = 4.5$ cm, and $L = 2.5\,\pi$ cm, with a representative $\kappa = 60$ W/mK. The 4 cm ($= R_o - R_i$) wide strip is divided into four 1 cm wide **regions** (Column 1), with an average $r$ (Column 2), and the use of Eqn. (S15) to compute the $k_r^c$ (Column 3) and $k_q^c$ (Column 4). Consequently, the relations in Eqn. S7b, were used to obtain $\kappa_1$ (Column 5) and $\kappa_2$ (Column 6).

| Region | Average "r" | $\kappa_r$ (W/mK) | $\kappa_\theta$ (W/mK) | $\kappa_1$ (W/mK) | $\kappa_2$ (W/mK) |
|---|---|---|---|---|---|
| 0.5 – 1.5 cm | 1 cm | 148 | 24 | 282 (= $a$) | 12 (= $b$) |
| 1.5 – 2.5 cm | 2 cm | 74 | 47 | 118 (= $c$) | 29 (= $d$) |
| 2.5 – 3.5 cm | 3 cm | 49 | 71 | 110 (= $e$) | 31 (= $f$) |
| 3.5 – 4.5 cm | 4 cm | 37 | 94 | 169 (= $g$) | 20 (= $h$) |